\documentclass[prx,twocolumn,superscriptaddress,nopacs]{revtex4}
\usepackage{graphicx}

\usepackage{amsmath}
\usepackage{amssymb}
\usepackage{accents}
\usepackage{color}
\usepackage{verbatim}
\usepackage{hyperref}
\usepackage{empheq} 
\definecolor{very-light-gray}{gray}{0.9}

\usepackage{framed}

\begin{document}

\title{Scaling Theory of Few-Particle Delocalization}
\author{Louk Rademaker}
\affiliation{Department of Theoretical Physics, University of Geneva, 1211 Geneva, Switzerland}

\date{\today}

\begin{abstract}
We develop a scaling theory of interaction-induced delocalization of few-particle states in disordered quantum systems. In the absence of interactions, all single-particle states are localized in $d<3$, while in $d \geq 3$ there is a critical disorder below which states are delocalized. We hypothesize that such a delocalization transition occurs for $n$-particle bound states in $d$ dimensions when $d+n\geq 4$. Exact calculations of disorder-averaged $n$-particle Greens functions support our hypothesis. In particular, we show that $3$-particle states in $d=1$ with nearest-neighbor repulsion will delocalize with $W_c \approx 1.4t$ and with localization length critical exponent $\nu = 1.5 \pm 0.3$. The delocalization transition can be understood by means of a mapping onto a non-interacting problem with symplectic symmetry. We discuss the importance of this result for many-body delocalization, and how few-body delocalization can be probed in cold atom experiments.
\end{abstract}

\maketitle


\section{Introduction}

The interplay between disorder and interactions in quantum systems is a long-standing open question in condensed matter physics. In the absence of interactions, single-particle wavefunctions are localized in $d=1,2$ dimensions, while in $d=3$ dimensions there is a transition from localized to delocalized states\cite{Anderson:1958fz,Abrahams:1979iv,MacKinnon:1981jt,1993RPPh...56.1469K,Evers:2008gi}. What happens in the presence of interactions is less clear. Though the existence of a {\em many-body localized} (MBL) phase is established in $d=1$ for strong disorder\cite{Basko:2006vz,Nandkishore:2015kt,Imbrie2016,Abanin:2019dl}, there is continuing debate what happens at the transition from ergodic to MBL, about the properties of the ergodic phase, and whether MBL can exist in higher dimensions $d>1$.

Here, we investigate the problem of interactions and disorder using the following perspective. Given that single-particle states are all localized in $d\leq2$ dimensions, how can many-particle states become {\em delocalized}?

It is, after all, highly unusual that a many-body state has fundamentally different elementary excitations with or without interactions. For example, within the widely applicable Fermi liquid theory\cite{Pines:1999wg} the single-particle excitations of the many-body state carry the same quantum numbers as the non-interacting states. In the case of many-body delocalization (MBdL), however, the elementary excitations of the many-body state are delocalized and thus have nothing in common with the single-particle fully localized states. Note that such a complete interaction-induced overhaul of the nature of excitations is also seen in the fractional Quantum Hall effect, Mott insulators or Luttinger liquids.

Often, however, the interaction-dominated many-body state can be understood in terms of bound states of a few particles. The most famous example is the Cooper pair instability of the Fermi liquid\cite{Cooper:1956el}. Similarly, the flux attachment argument for the $\nu=1/3$ FQHE shows how nontrivial many-body states can be reduced to having nontrivial few-body states. Therefore, we will look at the properties of $n$-particle states in interacting disorder quantum systems in $d$ dimensions.

The properties of {\em two-particle states} in $d=1$ dimensions have been studied extensively\cite{Shepelyansky:1994df,Imry:1995jq,Frahm:1995dj,Weinmann:1995gza,vonOppen:1996ff,Shepelyansky:1997di,Halfpap:2001br,Turek:2003iq}, showing that the localization length of two-particle bound states is enhanced (for weak disorder) compared to the single-particle localization length. Surprisingly, however, is that in $d=2$ dimensions an exact calculation of the two-particle Greens function showed a delocalization transition.\cite{Ortuno:1999vp,Cuevas:1999co} Three-particle states have been discussed using perturbation theory\cite{Shepelyansky:1997di,Xie:2012ud} and an increased localization length was observed for relatively large disorder\cite{Halfpap:2001br}, but no systematic results exist.

In fact, one can map interacting $n$-particle states onto the problem of noninteracting particles {\em with an internal structure}. For $d=2$, this relates the $n=2$-particle problem to the noninteracting case with symplectic symmetry, which can delocalize in $d=2$. Similarly, the internal structure of $n=3$-particle states in $d=1$ might be related to single particle problems that can delocalize in $d=1$.\cite{Evers:2008gi} Consequently, we propose the following hypothesis:
\begin{quote}
There exists a delocalization transition for $n$-particle bound states in $d$ dimensions when $n+d \geq 4$.
\end{quote}

A key insight that allows us to investigate this hypothesis is the existence of single-parameter {\em scaling}\cite{Abrahams:1979iv,1980PMagB..42..827A,MacKinnon:1981jt,MacKinnon:1983dn,Ortuno:1999vp}. This scaling implies that the $n$-particle localization length $\lambda_n (W,L)$ (for disorder strength $W$ and system size $L$) can be expressed as 
\begin{equation}
	\lambda_n(W,L) = L f_n^\pm (\lambda_n^\infty (W) / L)
	\label{Eq:Scaling1}
\end{equation}
where $\lambda_n^\infty (W)$ is the disorder-dependent localization length when the $n$-particle states are localized, or a scale of resistivity when the $n$-particle states are delocalized. The function $f_n^\pm$ only depends on the number of particles $n$, the dimension and whether it corresponds to a localized or delocalized branch.

In the remainder of this paper, we will first introduce the relevant model (Sec.~\ref{Sec:Model}), define $n$-particle Greens functions (Sec.~\ref{Sec:Greens}) and display numerical results (Sec.~\ref{Sec:Results}). 
In Sec.~\ref{Sec:Scaling} we will calculate the beta function of a renormalized $n$-particle transmission coefficient based on the scaling hypothesis.
To understand the numerical results, we will discuss a map from interacting particle systems onto noninteracting particle systems with symplectic symmetry in Sec.~\ref{Sec:SO}.
We conclude with a discussion of how our central hypothesis affects the properties of many-body localization in Sec.~\ref{Sec:MBL}.

\begin{figure*}
	\includegraphics[width=0.32\textwidth]{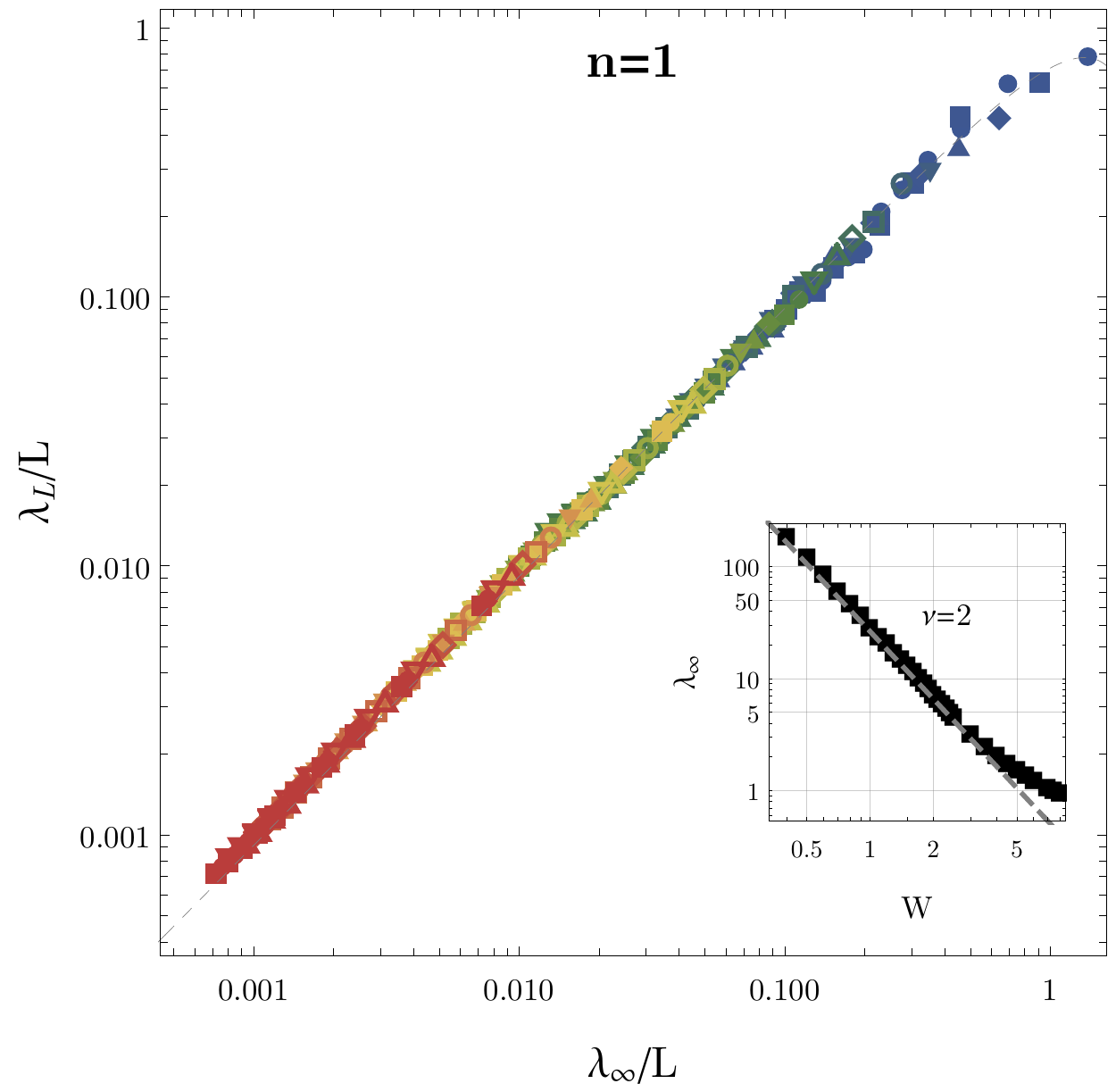}
	\includegraphics[width=0.32\textwidth]{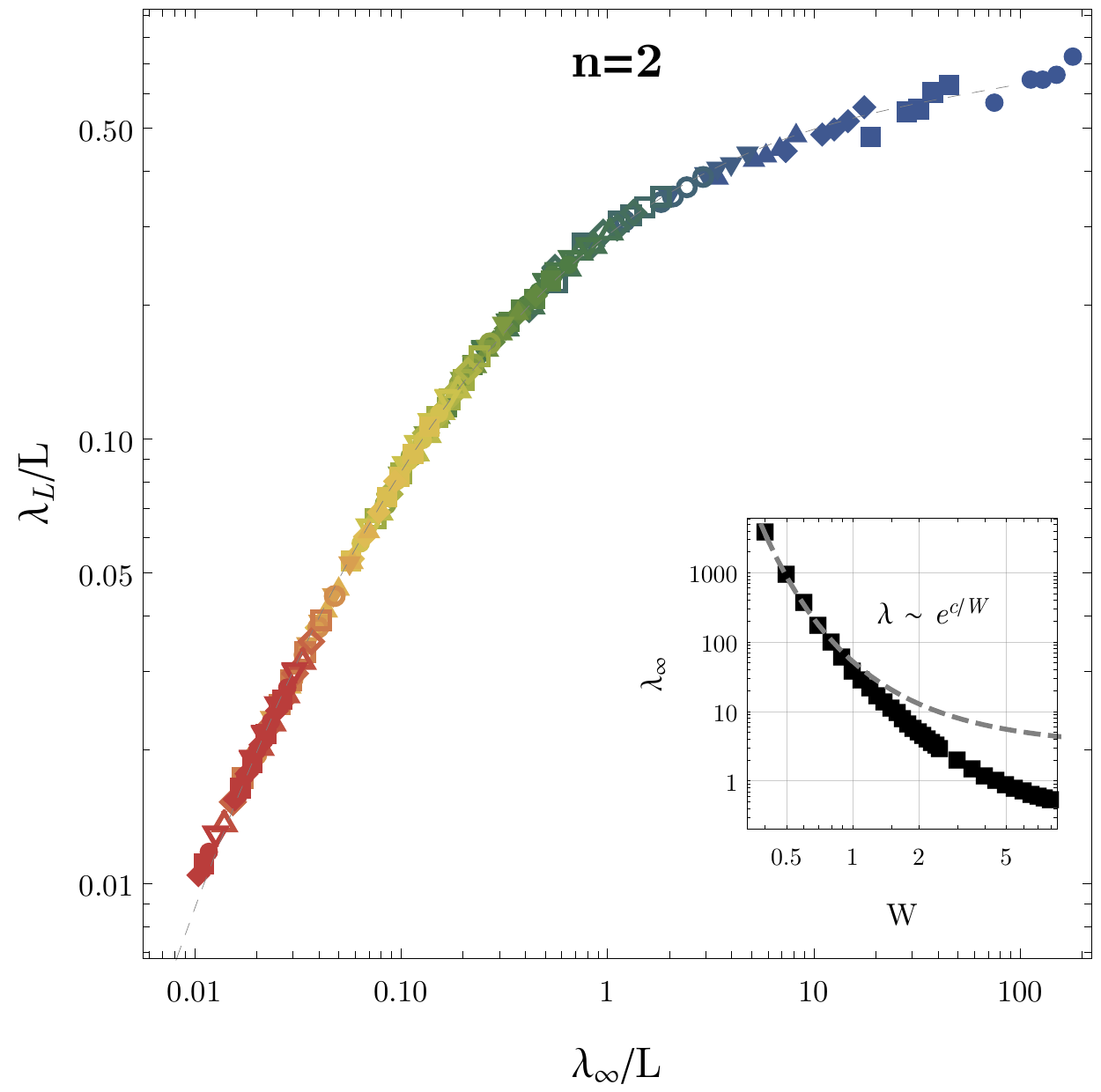}
	\includegraphics[width=0.32\textwidth]{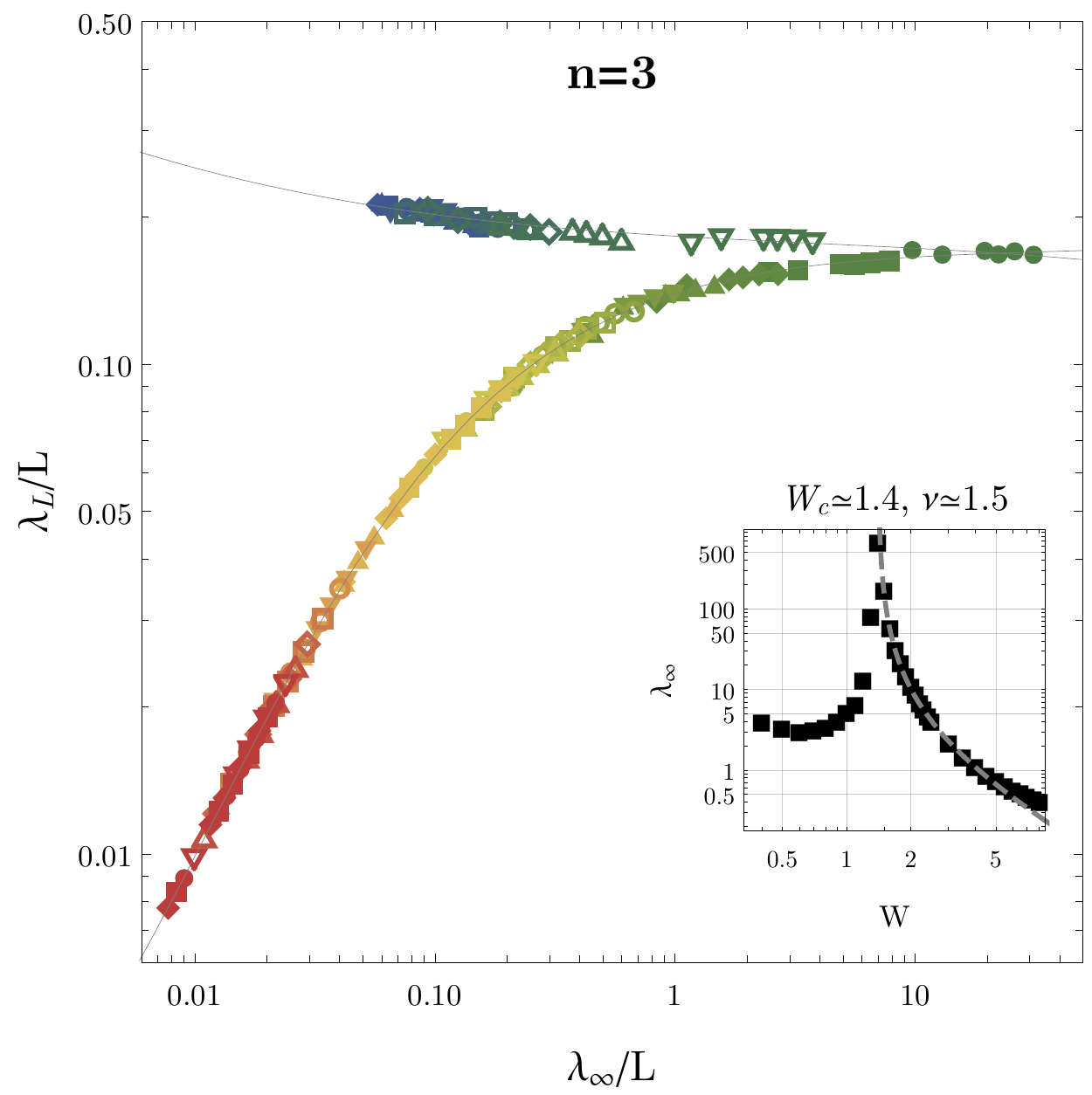} \\
	\includegraphics[width=0.98\textwidth]{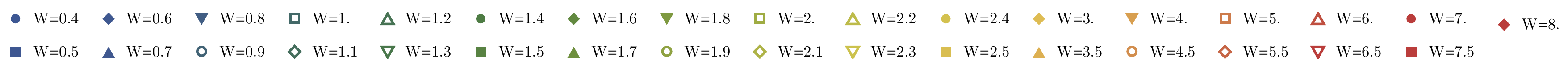}
	\caption{The scaling of the localization length $\lambda_n(W,L)$ for $n=1,2,3$-particle states. We display results up for $L=20, 24, 28, 32, 48$ for all disorder strengths, and up to $L=64$ for $1\leq W \leq 2$. For the single-particle states, we include data up to $L=1280$. The number of disorder realizations decreases from $N_d=2000$ for $L=20$ down to $N_d = 200$ for $L=64$ (for $n\geq 2$), for $n=1$ all data points have $N_d=2000$. The finite system size localization length $\lambda_n (W,L)$ is computed according to Eq.~\eqref{Eq:LLn}. The results for different system size and disorder strengths are then projected onto a single curve following the scaling ansatz Eq.~\eqref{Eq:ScalingF}. We clearly observe localization for $n=1,2$-particle states, with $\lambda^\infty_1 (W) \sim W^{-2}$ and $\lambda^\infty_2 (W) \sim e^{c/W}$. In contrast, the $n=3$-particle states undergo a delocalization transition at $W_c \approx 1.4$ with the localization length diverging as $\lambda^\infty_3(W) \sim |W-W_c|^\nu$ with exponent $\nu = 1.5 \pm 0.3$}
	\label{Fig:Numerics}
\end{figure*}

\section{Model}
\label{Sec:Model}

The 'standard model' of many-body localization is that of a chain of hopping spinless fermions with nearest-neighbor repulsion,\cite{Abanin:2019dl}
\begin{equation}
	H = - t \sum_{\langle i j \rangle} c^\dagger_i c_j 
	+ V \sum_{\langle i j \rangle} n_i n_j
	+ \sum_i \mu_i n_i
	\label{Eq:Hamiltonian}
\end{equation}
where $\mu_i$, the onsite chemical potential, is a random variable from a uniform distribution $\mu_i \in [ -W, W]$. We will consider chains with length $L$ and open boundary conditions. This model is equivalent, when $V=2t$, to the random field Heisenberg chain,\cite{Znidaric:2008cr,Pal:2010gr} which has been shown to exhibit a MBL-to-ergodic transition at a critical value $W_c \approx 3.6t$.\cite{Luitz:2015iv} In the absence of interactions, all single-particle eigenstates are localized, with a localization length of $\lambda_1 (W) = 105/4 (t/W)^2$ in the limit of small $W$.\cite{1993RPPh...56.1469K,Abrahams:1979iv,Anderson:1958fz} 

In the remainder of the paper we set $t=1$ and $V=2$.

\section{$n$-particle Greens functions}
\label{Sec:Greens}

Whether or not an $n$-particle state is delocalized or not, can be inferred from calculating the {\em Greens function}. For a single particle, the Greens function expresses the expectation value of creating a particle at position $y$ and retrieving it at position $x$. The Greens function over the full length of a chain with open boundary conditions thus corresponds to the likelihood of transmitting a particle through its entire length.

The single-particle Greens function is defined as
\begin{equation}
	G_1 (x;y; E) = \langle 0 | c_x (E-H)^{-1} c_y^\dagger | 0 \rangle
\end{equation}
which can be calculated using the {\em single-particle eigenstates} $\phi_{in}$ (such that $H^0_{ij} \phi_{jn} = \epsilon_n \phi_{in}$ where $H^0$ is real-space the single-particle Hamiltonian matrix),
\begin{equation}
	G_1 (x;y;E) = \sum_n \frac{\phi_{xn} \phi_{yn} }{E - \epsilon_n}.
\end{equation}
From now on, we always look at the middle of the spectrum, $E=0$. An exact calculation of the single-particle spectrum, and thus the single-particle Green's function, takes a computational time proportional to $\mathcal{O}(L^3)$.

For disordered systems defined by disorder strength $W$, we can disorder average the Greens function. In particular, for system with length $L$ the single-particle localization is defined as\cite{MacKinnon:1981jt}
\begin{equation}
	\lambda_1^{-1} (W, L) = - \frac{2}{L-1} \log \langle | G_1 (1;L)|^2 \rangle_{\mathrm{dis}}
	\label{Eq:LL1}
\end{equation}
This quantity can be related to the disorder-averaged single-particle transmission coefficient
\begin{equation}
	T_1 (W,L) = \exp \left( \frac{ -2 L }{\lambda_1 (W,L)} \right)
\end{equation}
For a localized system, the limit $\lim_{L \rightarrow \infty} \lambda_1 (W,L)$ yields the single-particle localization length. 

Similarly, one can define the $2$-particle Greens function
\begin{equation}
	G_2(x_1, x_2; y_1, y_2; E) = 
		\langle 0 | c_{x_2} c_{x_1} (E-H)^{-1} c_{y_1}^\dagger c_{y_2}^\dagger | 0 \rangle
	\label{Eq:GF2}
\end{equation}
In the absence of interactions, this 2-particle Greens function can be evaluated using the single-particle eigenstates, making use of the fact that fermions are indistinguishable particles that anticommute,
\begin{eqnarray}
	G_2^{(0)} 
	=\sum_{mn} \frac{ \phi_{x_2n} \phi_{x_1m} \phi_{y_1m} \phi_{y_2n}
		-\phi_{x_2m} \phi_{x_1n} \phi_{y_1m} \phi_{y_2n} }
	{E - \epsilon_m - \epsilon_n}
	\label{Eq:GF20}
\end{eqnarray}
The interacting 2-particle Greens function needs to include the effect of the interaction term $H_{\mathrm{int}} = V \sum_{\langle i j \rangle} n_i n_j$. An exact calculation of the 2-particle Greens function scales as $\mathcal{O}(L^6)$, because the 2-particle Hilbert space equals $\mathcal{O}(L^2)$. Following Von Oppen and co-workers,\cite{vonOppen:1996ff,Ortuno:1999vp} however, we can drastically speed up this calculation by using Dyson's equation.

Writing the 2-particle Greens function as a matrix in the $L(L-1)/2$-dimensionsal 2-particle Hilbert space, the Dyson equation reads
\begin{equation}
	G_2 = G_2^{(0)} + G_2^{(0)} H_{\mathrm{int}} G_2
	\label{Eq:2Dyson}
\end{equation}
where $G_2^{(0)}$ is the non-interacting Greens function from Eq.~\eqref{Eq:GF20}. The trick is to realize that the interactions are diagonal and only act on the $(L-1)$ states where the two particles are nearest neighbors. We can thus restrict the Dyson equation \eqref{Eq:2Dyson} to the subspace with neighboring particles,
\begin{equation}
	\tilde{G}_2 = \tilde{G}_2^{(0)} + \tilde{G}_2^{(0)} H_{\mathrm{int}} \tilde{G}_2
\end{equation}
where $\tilde{G}_2^{(0)}$ is the noninteracting 2-particle Greens function restricted to the states where the two particles are neighboring. Consequently, the interacting Greens function $\tilde{G}_2$ can be calculated quite efficiently: $\mathcal{O}(L^3)$ to get the single-particle eigenstates, $\mathcal{O}(L^4)$ to compute the noninteracting 2-particle Greens function in the restricted subspace, and $\mathcal{O}(L^3)$ to solve the Dyson equation. The leading contribution to the computing time thus comes from the calculation of the non-interacting Greens function, which still is much faster than a full exact diagonalization.

This efficient algorithm thus provides us the exact interacting 2-particle Greens function in the space where the two particles are always neighbors. Among them is the Greens function where we look at a particle pair going from sites $y_1,y_2=1,2$  to $x_1,x_2=L-1,L$, which amounts to a transmission of the particle pair through the length of the chain. This allows us to introduce, similar to Eq.~\eqref{Eq:LL1}, a 2-particle localization length
\begin{equation}
	\lambda_2^{-1} (W, L) = - \frac{2}{L-2} \log \langle | G_2 (1,2;L-1,L)|^2 \rangle_{\mathrm{dis}}
	\label{Eq:LL2}
\end{equation}

A similar efficient algorithm exists for the $n$-particle interacting Greens function, restricted to the subspace where the interactions are nonzero. The computational cost scales as $\mathcal{O}(L^{3n-2})$
and is always dominated by the calculation of the $n$-particle noninteracting Greens function. This is always more efficient than the $\mathcal{O}(L^{3n})$ needed for an exact diagonalization.

Given a general $n$-particle Greens function, the $n$-particle localization length is defined as
\begin{equation}
	\lambda_n^{-1} (W, L) = - \frac{2}{L-n} \log \langle | G_n(1\ldots n;L-n+1 \ldots L)|^2 \rangle_{\mathrm{dis}}
	\label{Eq:LLn}.
\end{equation}
This length can be related to the transmission probability of $n$-particle states through a chain of length $L$,
\begin{equation}
	T_n(W,L) = \exp \left( \frac{ -2 L }{\lambda_n (W,L)} \right).
	\label{Eq:TransmissionC}
\end{equation}


\section{Numerical results}
\label{Sec:Results}

We calculated the disorder-averaged $n$-particle localization length $\lambda_n(W,L)$ for $n=1,2,3$ in $d=1$ dimensions, for a range of system sizes from $L=20$ to $L=64$ (for $n=1$ we go up to $L=1280$). Following the one-parameter scaling theory of localization\cite{Halfpap:2001br,MacKinnon:1981jt,Abrahams:1979iv} we assume that the localization lengths for various sizes and disorder all collapse onto the same scaling function
\begin{equation}
	\lambda_n(W,L) / L = f^\pm_n (\lambda^\infty_n (W) / L)
	\label{Eq:ScalingF}
\end{equation}
where $\lambda^\infty_n$ is a fitting parameter, that depends only on the disorder strength $W$ and the particle number $n$. The function $f^\pm_n (x)$ only depends on the particle number $n$, and it can have a localized and delocalized branch.

The fitting parameter $\lambda^\infty_n$ is found as follows. For the strongest disorder, the system size is always bigger than the localization length and we set $\lambda^\infty_n = \lim_{L \rightarrow \infty} \lambda_n(W,L)$. The results for the next, smaller, disorder strength $W$ are then fitted such that we minimize the variance of $\log (\lambda_n(W,L)/L)$. Step by step, numerical results for smaller disorder values are included until a full smooth curve is obtained.

Our main numerical results are shown in Fig.~\ref{Fig:Numerics}. Consistent with earlier numerics and analytical arguments\cite{1993RPPh...56.1469K,MacKinnon:1983dn,MacKinnon:1981jt}, we find that the single-particle localization length indeed falls onto a curve that approaches $f_1(x) =x$ for large disorder. When $W$ is small, the localization length diverges as $\lambda_1^\infty = 105/4W^2$. The scaling curve $f_1(x)$ starts to deviate from $f_1(x)=x$ when the system size becomes of the order of the localization length, again consistent with the identification of $\lambda^\infty_1$ as the localization length of single particles.

We find that 2-particles states remain localized throughout, though the localization length is enhanced compared to the single-particle localization. This is consistent with earlier numerical and analytical work\cite{Halfpap:2001br,Shepelyansky:1997di,vonOppen:1996ff,Weinmann:1995gza,Frahm:1995dj,Imry:1995jq,Shepelyansky:1994df}, though the power with which $\lambda_2^\infty$ diverges is different from other works. In fact, our results are consistent with a singular divergence of the form $\lambda_2^\infty (W) \sim e^{c/W}$ where $c$ is some constant. The same divergence has also been observed for the single-particle localization length in $d=2$,\cite{MacKinnon:1981jt} strengthening the hypothesis that the $n=2$-particle bound states in $d=1$ correspond to $n=1$-particle states in $d=2$.

A new result is the appearance of a delocalization transition for $n=3$-particle states in $d=1$, as is shown in Fig.~\ref{Fig:Numerics}, right panel. For $W>W_c \approx 1.4t$ all states are localized. At low disorder, the results follow a {\em different scaling curve} than the localized curve, suggesting this upper branch represents delocalization. The localization length diverges, which is based on fitting the scaling curve, according to $\lambda^\infty \sim (W-W_c)^{-\nu}$ with exponent $\nu = 1.5 \pm 0.3$. The large uncertainty in the exponent is due to the uncertainty in the precise critical disorder strength. Note that this is in the same range as the critical exponent for the single-particle delocalization in $d=3$, namely $\nu_{d=3,n=1} = 1.2 \pm 0.3$\cite{MacKinnon:1981jt} and as $n=2$-particle delocalization in $d=2$, $\nu_{d=2,n=2}=1.2\pm 0.2$\cite{Cuevas:1999co}.

A final comment is in order with regard to the interpretation of $\lambda^\infty_n$. In the localized regime, this quantity corresponds to the localization length in an infinite system. When there is no localization, however, Ref.~\onlinecite{MacKinnon:1981jt} showed that the conductivity of the delocalized system is given by $\sigma (W) = 1/\lambda^\infty_{n=1} (W)$. As a generalization of their results, when $n \neq 1$, we interpret the inverse of $\lambda^\infty_n$ as an $n$-particle conductivity. It follows that the $n=3$-particle conductivity vanishes at the critical disorder strength $W_c$.


\section{Scaling theory}
\label{Sec:Scaling}

\begin{figure}
	\includegraphics[width=\columnwidth]{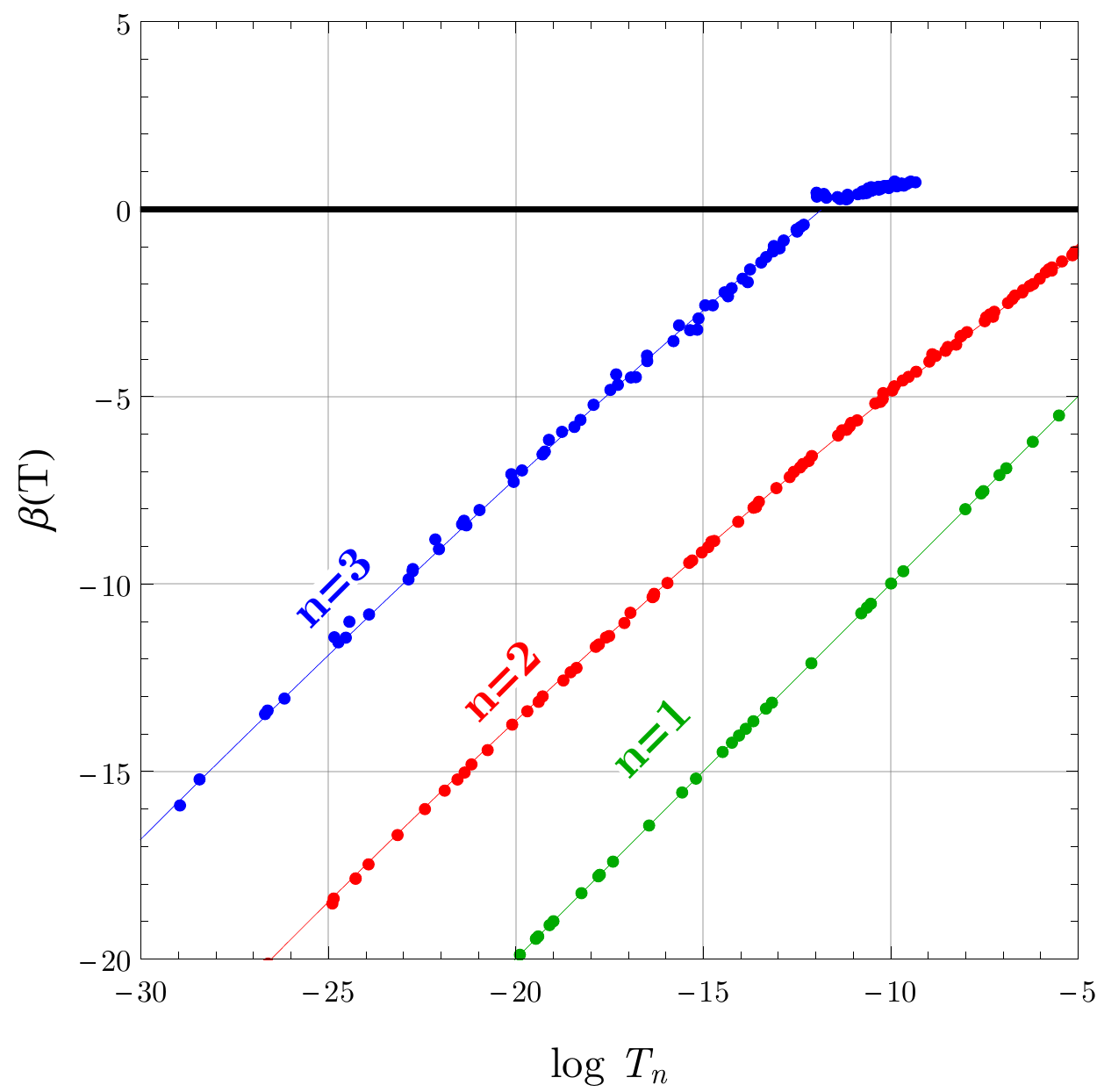}
	\caption{The scaling beta function $\beta(T)$ as a function of the transmission coefficient $T_n$ for $n=1,2,3$ particle states in $d=1$. The dots correspond to the data from Fig.~\ref{Fig:Numerics} scaled using Eq.~\eqref{Eq:BetaF}. This scaling diagram is similar to the single-particle scaling diagram in $d=1,2,3$ dimensions.\cite{MacKinnon:1981jt,Abrahams:1979iv}}
	\label{Fig:BetaF}
\end{figure}

The original scaling arguments of the 'Gang of Four' are based on {\em single-particle scaling},\cite{1980PMagB..42..827A,Abrahams:1979iv} meaning that the conductance $g$, that depends on the system size $L$, can be described by a scaling function
\begin{equation}
	\beta(g) \equiv \frac{ d \log g(L) }{ d\log L}
\end{equation}
The dimensionless conductance that is considered by the Gang of Four is related to the level spacing based on Thouless\cite{Thouless:1974ii}. However, conductance for few-particle states is not a well-defined concept.
Therefore, we consider instead the dimensionless {\em transmission coefficient} defined in Eq.~\eqref{Eq:TransmissionC}. Following the assumption of Eq.~\eqref{Eq:ScalingF}, we find
\begin{equation}
	\log T_n(W,L) = \frac{-2}{f_n (\lambda^\infty_n (W) / L) }.
\end{equation}
The corresponding $\beta$-function for this transmission coefficient can thus be extracted from the numerical data, in a method similar to the one employed by Ref.~\cite{MacKinnon:1981jt},
\begin{eqnarray}
	\beta(T) & = & \frac{ d \log T_n }{d \log L} \\
	& = & -2 \frac{ d( 1/ f_n (\lambda^\infty_n (W) / L) ) }{d \log L} \\
	& = & \frac{2}{f_n (\lambda^\infty_n(W) /L)} \frac{ d \log f_n (\lambda^\infty_n(W) /L ) }{d \log L} \\
	& = & \log T_n \, \frac{ d\log f_n (x)}{d \log x}
	\label{Eq:BetaF}
\end{eqnarray}
where in the last line we substituted $x = \lambda^\infty_n(W) /L$. As a consistency check, we know that for a localized system with $L \ll \lambda^\infty$, there are no finite size effects and thus $f(x) = x$, which results in $\beta (T) = \log T$ which implies $T = e^{-\alpha L}$.

We can verify the scaling using our numerical results, by explicitly calculating Eq.~\eqref{Eq:BetaF}. The results are shown in Fig.~\ref{Fig:BetaF}. The results are consistent with our hypothesis that $n=1,2,3$-particle states in $d=1$ have the same scaling behavior as single-particle states in $d=1,2,3$ dimensions. In addition with the earlier results that $n=2$-particle states in $d=2$ have scaling similar to single-particle states in $d=3$ dimension,\cite{Ortuno:1999vp} this further supports our hypothesis that localization scaling is determined by $n+d$ where $n$ is the number of particles and $d$ the dimensionality.

\section{Symplectic symmetry}
\label{Sec:SO}

The behavior of $n$-particle states in $d$ dimensions can be understood through a mapping onto single-particle problems with symplectic symmetry, also known as systems with spin-orbit coupling. 

Using diagrammatic expansions and renormalization group analysis, Hikami et al.\cite{Hikami:1980jn} showed that the presence of spin-orbit coupling changes the sign of the weak localization. This is due to the existence of an anti-unitary symmetry $T$ that squares to $T^2=-1$, in other words, the existence of a {\em symplectic symmetry}.\cite{Hikami:1980jn,Efetov:1983hm} In the case of $d=2$ dimensions, spin-orbit coupling therefore allows for a delocalization transition of single-particle states. This has been confirmed both experimentally, as seen in the magnetoresistance\cite{Bergman:1982fn,Bergmann:bf1984}, as well as in extensive analytical\cite{Efetov:1983hm} and numerical studies\cite{mackinnon1985scaling,Schweitzer:1997et,Evangelou:1995cf,Ando:1989if,Asada:2002it,Evangelou:1987hy,Merkt:1998ca}. 

While symplectic delocalization is widely studied in $d=2$, also in $d=1$ the symplectic symmetry class (AII, which requires just a symmetry $T$ with $T^2=-1$) can give rise to exactly one delocalized state provided there are an odd number of channels.\cite{Evers:2008gi,Zirnbauer:1992dc,Ando:2002es,Mirlin:1994hj} In this section, we will aim to interpret the interacting cases $n=2,d=2$ and $n=3,d=1$ in terms of non-interacting particles with an internal structure with approximate symplectic symmetry. 

In general, the presence of an internal degree of freedom for noninteracting particles allows for generic random hopping
\begin{equation}
	H' = \sum_{\langle ij \rangle, \alpha \beta}
	V^{\langle ij \rangle}_{\alpha \beta}
	( b^\dagger_{i \alpha} b_{j \beta} + h.c. )
	\label{Eq:HamiltonianSymplectic}
\end{equation}
where $b^\dagger_{i \alpha}$ creates a particle at site $i$ with internal state $\alpha$. In the case of $d=2$-dimensional symplectic symmetry, the hopping matrix $V^{\langle ij \rangle}_{\alpha \beta}$ can be parametrized using Pauli matrices $\sigma^a$,
\begin{equation}
	V^{\langle ij \rangle}_{\alpha \beta} = \sum_x V^{\langle ij \rangle}_a \sigma_{\alpha \beta}^a
	\label{Eq:2-stateSymmetry}
\end{equation}
where $V^{\langle ij \rangle}_a$ are local independent random variables. The Pauli matrices $\sigma^a$ act on the internal structure of the particles. This structure was studied numerically in $d=2$ in Refs.~\onlinecite{mackinnon1985scaling,Evangelou:1995cf,Ando:1989if,Evangelou:1987hy}.

The hopping matrix of Eq.~\eqref{Eq:2-stateSymmetry} has symplectic symmetry in the form of the operator $T = \sigma^y K$ where $K$ is charge conjugation. Here $T$ acts as an anti-unitary symmetry of the hopping terms. However, instead of $\sigma^y$ acting on a 'real' spin degree of freedom, it acts on the internal state of the $b$-particles. Nevertheless, the presence of a $T^2=-1$ symmetry places this effective model in the AII symmetry class.

Let us now see what the possible internal states are for interacting few-body states in $d=2$ and $d=1$ dimensions.

\subsection{Two dimensions}

To study two-particle states in $d=2$ dimensions, consider the most general two-particle state $|\psi \rangle = \sum_{ij} \phi_{ij} c^\dagger_i c^\dagger_j | 0 \rangle$, with $\phi_{ij}$ an antisymmetric wavefunction. Rather than expressing the wavefunction in terms of the two positions $i,j$, it is more natural to change to a basis with the center-of-mass $R = i+j$ and the relative position $r=i-j$. The short-range interaction only depends on the relative coordinate, $V(r)$. Therefore, the wavefunction for fixed $R$ is described by a quantum-mechanical problem of a particle in the potential $V(r)$. Such a problem can be decomposed in states with two-dimensional angular momentum $\ell$ and a radial profile characterized by quantum number $n$. The wavefunction of the two-particle state is therefore $\phi_{n\ell} (R)$. This wavefunction is now viewed as a single-particle wavefunction for the center-of-mass coordinate $R$ with an internal structure characterized by the quantum numbers $n \ell$. The presence of disorder induces scattering between the internal states depending on the exchanged momentum, which is analogous to the random hopping Hamiltonian of Eq.~\ref{Eq:HamiltonianSymplectic}. This establishes on a formal level the equivalence of the problem of symplectic non-interacting particles with $n=2$-particle states in $d=2$. Since non-interacting symplectic particles in the AII class can delocalize, so can the $n=2$-particle states. Note that within the numerical accuracy, the critical exponent of both classes is the same.\cite{mackinnon1985scaling,Ortuno:1999vp}

We can make this mapping explicit using strong-coupling perturbation theory. Of course, we are not actually in the limit of strong coupling. However, this method allows us to systematically construct an effective single-particle Hamiltonian. The intermediate-coupling regime (in our numerical results we have $V=2t$) can be reached by higher-order expansions. Nevertheless, the effective symmetry of the resulting Hamiltonian is already reflected at the 2nd order in perturbation theory.

When $V \gg t$ in Eq.~\ref{Eq:Hamiltonian} in $d=2$, there are two inequivalent interacting states: having the particles neighbor each other along an $x$ or $y$ bond of the square lattice. Let us call a state $| R \alpha \rangle = c^\dagger_R c^\dagger_{R + \alpha} |0 \rangle$ where $\alpha = x,y$. In second-order perturbation theory, the effective hopping Hamiltonian for these states has the following nonzero components,
\begin{eqnarray}
	\langle R' \alpha' | H_{\mathrm{eff}} | R \alpha \rangle & = & 
	\frac{t^2}{V} \left( \delta_{\alpha \alpha'} 
	( \delta_{|R-R'| = \alpha} + 2 \delta_{|R - R'| = \overline{\alpha}} )
	\right. \nonumber \\ && \left.
	+ 2 \delta_{\overline{\alpha} \alpha'} 
	( \delta_{RR'} + \delta_{R-R'=\alpha'}
	\right. \nonumber  \\ && \left.
	+ \delta_{R'=R+\alpha} + \delta_{R'=R+\alpha-\alpha'})
	\right).
\end{eqnarray}
We use the notation here that $\overline{x} = y$ and $\overline{y}=x$. The above expression is quite tedious, but it can be simplified by an appropriate symmetrization of our basis set of states. To this end, introduce the following set of states,
\begin{equation}
	| \psi_{R \pm} \rangle =
		\frac{1}{2} \left[ ( | R x \rangle + |(R-x) x \rangle ) \pm 
		( | R y \rangle + |(R-y) y \rangle )\right] .
\end{equation}
These states have center-of-mass $R$ and carry angular momentum eigenvalue $\ell = \pm 1$. Note that this exhausts the possible values of the angular momentum for the $C_4$ rotational symmetry of the square lattice. The nearest-neighbor hopping of these states is given by
\begin{equation}
	\langle \psi_{R+\delta \sigma'} | H_{\mathrm{eff}} | \psi_{R \sigma} \rangle
	= \frac{t^2}{2V} \left( 3 + 4 \sigma \sigma' \right).
	\label{Eq:d2SymplecticHopping}
\end{equation}
where $\delta = \pm x,y$ is the nearest-neighbor vector and $\sigma,\sigma'$ are $\pm1$. Note that all hoppings are real and symmetric in $\sigma, \sigma'$. Consequently, the anti-unitary operator $T = \sigma^y K$ is a symmetry of this effective hopping model.

At this order, the onsite disorder affects the states $|\psi_{R \sigma} \rangle$ independent of $\sigma$. At higher order in perturbation theory, the mix of disorder and hopping can give rise to random hoppings. Nevertheless, in this systematic expansion the symplectic symmetry of Eq.~\eqref{Eq:d2SymplecticHopping} remains intact. There is a major caveat to this line of reasoning: the perturbation theory might break down before we reach the intermediate coupling regime. Here is where the numerics come in: based on Refs.~\onlinecite{Ortuno:1999vp,Cuevas:1999co} we estimate that the delocalizated regime also exists for intermediate interaction strengths.

\subsection{One dimension}

In $d=1$, two-particle states do not have internal structure in general. This is clearly seen in our model with nearest-neighbor repulsion. The only possible interacting 2-particle states are living on a neighboring bond, $| \psi_2 (R) \rangle  = c^\dagger_{R} c^\dagger_{R+1} | 0 \rangle$. While the two-particle states do feel (on average) a weakened disorder-profile due to the extent of its wavefunction, this only leads to an increased localization length and not to delocalization.\cite{Shepelyansky:1994df,Imry:1995jq,Frahm:1995dj,Weinmann:1995gza,vonOppen:1996ff,Shepelyansky:1997di,Halfpap:2001br,Turek:2003iq}

However, $n=3$ appears to be the smallest possible number of particle required to obtain an effective single-particle system with a nontrivial internal structure. Consider the three states $t^\dagger_{R\sigma}$ defined by
\begin{eqnarray}
	b^\dagger_{R0} & = & c^\dagger_{R-1} c^\dagger_{R} c^\dagger_{R+1} , \\
	b^\dagger_{R\pm} & = & \frac{1}{\sqrt{2}} c^\dagger_R (c^\dagger_{R-1} c^\dagger_{R+2} 
		\pm c^\dagger_{R-2} c^\dagger_{R+1} ) .
\end{eqnarray}
The original hopping Hamiltonian restricted to the space with $t_{R\sigma}$ states reads
\begin{eqnarray}
	H_{\mathrm{eff}} &=& \sum_R b^\dagger_{R \alpha}( - t S^x_{\alpha \beta} + V (S^z)^2_{\alpha \beta}) b_{R \beta} \nonumber \\
	&& + \sum_{R} t^\dagger_{R \alpha}(-t \lambda_4)_{\alpha \beta} t_{R +1, \beta} + h.c.
\end{eqnarray}
where $S^\alpha$ are spin-1 operators and $\lambda_4 = {\tiny \begin{pmatrix} 0 & 0 & 1 \\ 0 & 0 & 0 \\ 1 & 0 & 0 \end{pmatrix}}$.

These are the first three states of a tower of three-particle states that are either symmetric or antisymmetric with respect to reflection symmetry around the middle site $R$. The regular spin-1 time-reversal symmetry $T = e^{-i \pi S_y} K$ squares to $T^2=+1$, so this does not provide us with a symplectic symmetry. However, the hopping term is symmetric under the anti-unitary operation $T' = \lambda_5 K$ with $\lambda_5 = {\tiny \begin{pmatrix} 0 & 0 & -i \\ 0 & 0 & 0 \\ i & 0 & 0 \end{pmatrix}}$ that exchanges the $\sigma = \pm$ states. This operation only acts on part of the Hilbert space, but in there it acts as $T'^2=-1$. 

At the moment it is unclear to us whether this places the effective non-interacting Hamiltonian in the symplectic class AII. However, if it would, this would explain the delocalization in terms of AII systems with an odd number of channels.\cite{Evers:2008gi,Zirnbauer:1992dc,Ando:2002es,Mirlin:1994hj}

\section{Many-body delocalization}
\label{Sec:MBL}

Up till now we have shown how few-particle bound states, in particular $n=3$-particle states in $d=1$ and $n=2$-particle states in $d=2$, can delocalize in a model with interactions and disorder. It is an open question to what extent these delocalized bound states affect the many-body state where the number of particles scales $n$ with the system size $L^d$.

Early on, Imry\cite{Imry:1995jq} already speculated that a "serious rearrangement" will occur when single-particle states are localized but some few-particle states are delocalized. In particular, transport properties of finite size system with finite particle density might pick up the fact that not single-particle but only $n=2,3$-particle states can traverse the whole system. It was argued that many-body delocalization occurs "as soon as finite but mobile excitations exist"\cite{deRoeck:2016hya}. If indeed delocalization is driven by $n$-particle mobile excitations, this can be observed in the form of $h/ne$ Aharonov-Bohm oscillations.\cite{Weinmann:1995gza} Similarly, whether only $n$-particle states contribute to the mesoscopic transport can be probed using noise spectroscopy.\cite{Landauer:1998jh} 

It might be interesting to seek such effects in cold atoms systems, which are known to be excellent simulators of many-body localization.\cite{Luschen:2017ed,Schreiber:2015jta} The properties of few-particle delocalized states can be either studied in systems with explicitly $n$ particles, or in many-body systems at finite particle density.

Our simulations indicate that the $3$-particle delocalization transition occurs at $W^{n=3}_c \approx 1.4$, whereas the MBL transition occurs at a higher critical disorder $W_c^{\mathrm{MBL}} \approx 3.6$.\cite{Znidaric:2008cr,Pal:2010gr,Luitz:2015iv} This means for some part of the ergodic phase even $3$-particle states are localized. Because $4$-particle states will be more delocalized than $3$-particle states, we expect the critical disorder for delocalization to be higher, $W^{n=4}_c > W_c^{n=3}$. This generalizes to $W_c^n > W_c^{n-1}$. Naturally all $n$-particle states must be localized in a full MBL state, which provides the limit $W_c^n < W_c^{\mathrm{MBL}}$. There must be therefore a sequence of $n$-particle delocalization transitions with $\lim_{n \rightarrow \infty} W_c^n = W_c^{\mathrm{MBL}}$ in the thermodynamic limit. 

For a given disorder strength in the ergodic regime, there is thus a critical number $n^*(W)$ of particles that are needed in order for delocalization. Clusters of particles with $n>n^*$ can diffuse, whereas isolated particle clusters with $n<n^*$ will remain localized. In the dynamics of such a system, the delocalized particle states can collide with localized ones, thereby either breaking up into smaller (more localized) clusters or combining into larger (more delocalized) clusters. From a transport perspective, the dynamics can thus be described by a combination of random walks, localization, and transitions between the two. As a result, the larger $n^*$ the more rare the $n>n^*$ clusters become, and the more rare the periods of diffusion become. This might explain the subdiffusive ergodic regime, not in terms of rare regions ('Griffiths effects'), but in terms of larger and larger clusters of particles that are needed to delocalize.\cite{Luitz:2017cp} Similarly, the fact that only large clusters of particles contribute to ergodic behavior might explain why there are no signatures of ergodicity in average properties of integrals of motion.\cite{Rademaker:2017ky}

The existence of an 'ergodic bubbles' has been argued to prevent many-body localization, in particular in higher dimensions $d>1$\cite{deRoeck:2017ia,Luitz:2017ia}. However, such bubbles were proposed due to local disorder fluctuations - akin to Griffiths regions - instead of few-particle bound states studied here. The same mechanism, however, could still lead to the absence of full MBL in $d=2$. In this scenario, the critical disorder for $n$-particle delocalization diverges with large $n$. As a consequence, all many-body states in $d=2$ are delocalized but will exhibit subdiffusive behavior for a wide range of disorder strengths. This might explain the experimentally observed finite-time localization in $d=2$ cold atom systems.\cite{2016Sci...352.1547C}

Finally, we speculate about the shape of the ergodic many-body states for low disorder. Just like a superconducting state can be made out of Cooper pairs, an eigenstates in the many-body ergodic regime could be constructed out of a finite density of delocalized $3$-particle bound states. 
We leave an analysis of interactions between such $3$-particle states, and how to combine many of them in a variational many-body wavefunction, for future research.

Note that many-body localization induced by mobile particles was also recently discussed in Ref.~\onlinecite{Krause:2021il}.

\section{Outlook}

We introduced the hypothesis that $n$-particle states in $d$-dimensional interacting disordered systems have a delocalization transition when $n+d \geq 4$. Our numerical results for $n=1,2,3$ and $d=1$, together with earlier results in $d>1$, seem to confirm this hypothesis. The transmission of $n$-particle bound states can be captured within a single-parameter scaling theory. We further analyse the delocalization in terms of a mapping to symplectic non-interacting problems.

In our model, Eq.~\eqref{Eq:Hamiltonian}, we included spinless fermions and short-range interactions. It is known, however, that localization is affected by the presence of spin-orbit coupling or long-range interactions (such as the realistic Coulomb interaction).\cite{Rademaker:2020fo}  The study of the role of few-body delocalization in models that include such effects is still an open question.

The main difficulty with the scaling theory presented here is the limited power of exact numerical solutions, which are at its heart. Luckily, recent developments in cold atoms experiments have been able to successfully reproduce predictions related to many-body localization. We hope that the role of few-body delocalization can be similarly confirmed experimentally.

\begin{acknowledgments}
The author thanks Miguel Ortu\~{n}o, Andres Somoza, Dima Abanin, Vladimir Dobrosavljevic, Michael Sonner, Michele Filippone, Thierry Giamarchi and Markus M\"{u}ller for useful discussions and comments. 
The author was supported by Ambizione Grant No. PZ00P2\_174208/1 from the Swiss National Science Foundation.
\end{acknowledgments}


%

\end{document}